\documentclass[12pt]{article}

\usepackage{cite}

\usepackage{times}
\usepackage{graphicx}
\usepackage{journals}

\newenvironment{sciabstract}{%
\begin{quote} \bf}
{\end{quote}}

\newcounter{lastnote}
\newenvironment{scilastnote}{%
\setcounter{lastnote}{\value{enumiv}}%
\addtocounter{lastnote}{+1}%
\begin{list}%
{\arabic{lastnote}.}
{\setlength{\leftmargin}{.22in}}
{\setlength{\labelsep}{.5em}}}
{\end{list}}

\title{The Geometric Distance and Proper Motion of the Triangulum
  Galaxy (M33)}

\author
{Andreas Brunthaler,$^{1,2}$ Mark J. Reid,$^{3}$ Heino Falcke,$^{4,5}$\\
  Lincoln J. Greenhill,$^{3}$ Christian Henkel$^{1}$\\
\\
\normalsize{$^{1}$Max-Planck-Institut f\"ur Radioastronomie, Auf dem H\"ugel 69,
53121 Bonn, Germany}\\
\normalsize{$^{2}$Joint Institute for VLBI in Europe, Postbus 2, 7990
  AA Dwingeloo, The Netherlands}\\
\normalsize{$^{3}$Harvard-Smithsonian Center for Astrophysics, 60
  Garden Street, Cambridge, MA 02138, USA}\\
\normalsize{$^{4}$ASTRON, Postbus 2, 7990 AA Dwingeloo, The
  Netherlands}\\
\normalsize{$^{5}$Department of Astrophysics, Radboud Universiteit
  Nijmegen,}\\
\normalsize{Postbus 9010, 6500 GL Nijmegen, The Netherlands}\\
}

\date{}

\begin{document} 


\baselineskip24pt


\maketitle


\begin{sciabstract}
We measured the angular rotation and proper motion of the Triangulum
Galaxy (M33) with the Very Long Baseline Array by observing two H$_2$O
masers on opposite sides of the galaxy. By comparing the angular
rotation rate with the inclination and rotation speed, we obtained a
distance of 730 $\pm$ 168 kiloparsecs. This distance is consistent with
the most recent Cepheid distance measurement. M33 is moving with a
velocity of 190 $\pm$ 59 km~s$^{-1}$ relative to the Milky Way. These
measurements promise a new method to determine dynamical models for the
Local Group and the mass and dark matter halos of M31, M33 and the
Milky Way.    
\end{sciabstract}



Measuring the proper motion and geometric distances of nearby galaxies
has been a long-standing problem. As part of a famous debate about the
nature of galaxies, van Maanen -- an experienced observer --- claimed 
in 1923 to have measured a large proper motion and angular rotation rate
for the Triangulum Galaxy (M33) on photographic plates separated by 12 years
\cite{vanMaanen1923}. This was proven wrong by Hubble through the
discovery of Cepheids in M33 that showed a large distance
\cite{Hubble1926}. This pushed the detection of galaxy proper motions
beyond the capabilities of past experiments. Yet, galaxy proper motions
are important for many astrophysical issues, of which two are addressed
in this report. 


First, measuring accurate distances is of great importance to all
fields of astrophysics, from stellar astronomy to cosmology. The
calibration of most standard candles used for measuring extragalactic
distances is tied directly or indirectly to the distance to one galaxy,
the Large Magellanic Cloud (LMC), which remains controversial
\cite{FreedmanMadoreGibson2001,UdalskiPietrzynskiWozniak1998}. Hence,
it is important to obtain geometric distances to nearby galaxies in
which well understood standard candles can be studied. This allows
independent calibration and verification of the extragalactic distance
scale.

Another important issue is the distribution of luminous and dark matter
in the local universe. The problem when trying to derive the
gravitational potential of the Local Group of galaxies \cite{LG} is
that usually only radial velocities are known from the Doppler effect
and statistical approaches have to be used
\cite{HartwickSargent1978,KulessaLynden-Bell1992}.    
The proper motions of some nearby galaxies in the Milky Way subgroup
have been obtained from comparing historic photographic plates
\cite{JonesKlemolaLin1994,SchweitzerCudworthMajewski1995}, but a
confirmation of these measurements will require decades.
With Very Long Baseline Interferometry (VLBI) and phase-referencing
techniques, the proper motions for galaxies within the Local Group,
on scales of tens of microarcseconds per year, can now be measured.  The
most suitable strong and compact radio sources in the Local Group for
such a VLBI  experiment are the strong H$_2$O masers in M33 and in the
galaxy IC\,10 \cite{ArgonGreenhillMoran2004,ArgonGreenhillMoran1994}.


We observed two H\,II regions \cite{IsraelvanderKruit1974} in M33 with
known H$_2$O maser activity -- M33/19 and IC\,133 --
with the NRAO Very Long Baseline Array (VLBA) eight times between March
2001 and January 2004. M33/19 is located in the south eastern part of
M33, whereas IC\,133 is located in the north east of M33 ({\it
  Fig.1}). Our observations are grouped into four epochs, each
comprising two closely spaced observations to enable assessment of
overall accuracy and systematic errors (Table~\ref{obsinfo}). The
separations of the two observations within each epoch were large enough
that the weather conditions were uncorrelated, but small enough that
changes in the positions were negligible during this time.

\begin{table}
      \caption[]{Details of the observations: Observing date,
      observation length $t_{obs}$, beam size $\theta$ and position
      angle $PA$.} 
         \label{obsinfo}
      \[
  \begin{tabular}{p{0.08\linewidth}cp{0.10\linewidth}cp{0.10\linewidth}cp{0.10\linewidth}cp{0.10\linewidth}cp{0.08\linewidth}}
           \hline
Epoch& Date & $t_{obs} [h]$  & $\theta$ [mas] &$PA [^\circ]$\\
            \hline
     I&      2001/03/27 & 10  &0.88$\times$0.41 &164 \\
     I&      2001/04/05 & 10  &0.86$\times$0.39 &169 \\
            \hline
     II&      2002/01/28 & 10  &0.62$\times$0.33 &176\\
     II&      2002/02/03 & 10  &0.71$\times$0.33 &175 \\
            \hline
     III&      2002/10/30 & 10  &0.87$\times$0.38 &171 \\
     III&      2002/11/12 & 10  &0.84$\times$0.36 &165 \\
            \hline
      IV&     2003/12/14 & 12  &0.85$\times$0.36 &159 \\
      IV&     2004/01/08 & 12  &1.15$\times$0.47 &164 \\
            \hline
         \end{tabular}
      \]
   \end{table}

We observed four 8 MHz bands, in dual circular
polarization. The 128 spectral channels in each band yielded a channel
spacing of 62.5 kHz, equivalent to 0.84 km s$^{-1}$, and covered a
velocity range of 107 km s$^{-1}$.  The observations involved rapid
switching between the phase-calibrator J0137+312, which is a compact
background source with continuum emission, and the target sources
IC\,133 and  M33/19 in the sequence J0137+312 -- IC\,133 -- J0137+312
-- M33/19 -- J0137+312. With source changes every 30 seconds, an
integration time of 22 seconds was achieved.  The background source was
unresolved in all epochs and was assumed to be stationary on the
sky. Because J0137+312 was separated by only 1$^\circ$ on the sky from
the masers, we obtained a precise angular separation measurement for
all sources.   

The data were edited and calibrated with standard techniques
in the Astronomical Image Processing System (AIPS) as well as zenith
delay corrections (\citen{ReidBrunthaler2004}).
The masers in IC\,133 and M33/19 were imaged with standard techniques
in AIPS. In IC\,133, we detected 29 distinct emission features in
position and the spatial distribution was very similar to earlier observations
\cite{ArgonGreenhillMoran2004,GreenhillMoranReid1990,GreenhillMoranReid1993}.
All components were unresolved. In M33/19, we detected eight maser
features. Two features were separated by less than a beam size and
blended together. These two features were fit by two elliptical Gaussian
components simultaneously. All other features were fit by a single
elliptical Gaussian component.


The maser emission in M33/19 and IC\,133 is variable on time scales
less than 1 year. Between the epochs, new maser features appeared and
others disappeared. However, we were able to detect and
follow the motions of four features in M33/19 and six features in
IC\,133 over all four epochs. The feature identification was based
on the positions and radial velocities of the maser emission. Each
feature was usually detected in several frequency channels. A
rectilinear motion was fit to each maser feature in each velocity
channel separately. We discarded fits with reduced $\chi^2$ larger
than 3 as they are likely affected by blending or component
misidentification. All features showed consistent motions within their
errors (2$\sigma$). The accuracy and number of measured motions was not
adequate to model the internal dynamics of the IC133 and M33/19 regions
(such as outflow) as was done in earlier observations
\cite{GreenhillMoranReid1993,ArgonGreenhillMoran2004}.   

We then calculated the variance weighted average of all motions.  This
yielded an average motion of the maser features in M33/19 of 35.5 $\pm$
2.7 $\mu$as year$^{-1}$ in right ascension and $-$12.5 $\pm$ 6.3 $\mu$as
year$^{-1}$ in declination relative to the background source. For IC\,133
we get an average motion of 4.7 $\pm$ 3.2 $\mu$as year$^{-1}$ in right
ascension and $-$14.1 $\pm$ 6.4 $\mu$as year$^{-1}$ in declination. 


We also calculated the average position of all maser features
for each observation ({\it Fig.~2 and~3}). We used the individual
fits for each maser feature to remove a constant position offset
for each maser feature. We used the position offsets at the time
2002.627, which is in the middle of our observations. We then
calculated the variance weighted average of the positions of all
detected features. A fit of a rectilinear motion to these average
positions yielded motions of 37 $\pm$ 5 $\mu$as year$^{-1}$ in right
ascension and $-$13 $\pm$ 6 $\mu$as year$^{-1}$ in declination for
M33/19. For IC\,133 we obtained a motion of 3 $\pm$ 3  $\mu$as year$^{-1}$ in
right ascension and $-$13 $\pm$ 10 $\mu$as year$^{-1}$ in
declination. This is consistent with the variance weighted average of
all maser feature motions and suggests that the systematic internal
motions within the two regions (such as outflow) are probably not a
substantial source of bias.  The difference in the maser position of
the two closely spaced observations within each epoch was used to
estimate the accuracy of the position measurements. The average
position error over the four epochs was 7.7 $\mu$as in right ascension
and 9.8 $\mu$as in declination.


The relative motions between M33/19 and IC\,133 are independent of the
proper motion of M33 and any contribution from the rotation of the
Milky Way. Knowing the rotation curve and inclination of the galactic
disk we can predict the relative angular motion of
the two masing regions as a function of distance. The rotation of the
H\,I gas in M33 has been measured and the measured velocities were fit
with a tilted-ring model \cite{CorbelliSchneider1997}.  We
used this model of the rotation of M33  to calculate the expected
transverse velocities of M33/19 and IC\,133. For M33/19, we expect a
motion of 42.4 km~s$^{-1}$ in right ascension and $-$39.6 km~s$^{-1}$
in declination. For IC\,133 we expect $-$64.0 km~s$^{-1}$ in right
ascension and $-$74.6 km~s$^{-1}$ in declination. This gives a relative
motion of 106.4 km~s$^{-1}$ in right ascension and 35 km~s$^{-1}$ in
declination between the two regions of maser activity. 

The radial velocities of the H$_2$O masers in M33/19 and IC\,133 and
the H\,I gas at the same positions agree ($<$ 10 km~s$^{-1}$). This
suggests that the maser sources are moving with the H\,I
gas in the galaxy. However, although the rotation model and the radial
velocity of the H\,I gas at the position of IC\,133 is also consistent
($<$ 5 km~s$^{-1}$), there is a difference of $\sim$ 15 km~s$^{-1}$ at
the position of M33/19. This indicates the presence of another tilt in
the disk that is not covered by the model. Because of this uncertainty
in the rotation model, we conservatively assume a systematic error of
20 km~s$^{-1}$ in each velocity component for the relative velocity of
the two maser components.   

Comparing the measured angular motion of 30.8 $\pm$ 4 $\mu$as year$^{-1}$
in right ascension with the expected linear motion of 106 $\pm$ 20
km~s$^{-1}$, one gets a geometric distance of 

\begin{eqnarray}
\nonumber
D=730 \pm 100 \pm 135~\mathrm{kpc}, 
\end{eqnarray}
where the first error indicates the statistical error from the proper
motion measurements while the second error is the systematic error from
the rotation model. After less than three years of observations, the
uncertainty in the distance estimate is dominated by the uncertainty of
the rotation model of M33. 


Within the errors the geometric distance of $730 \pm 100 \pm
135~\mathrm{kpc}$ is consistent with recent Cepheid and tip of
the red giant branch (TRGB) distances of 802$\pm$51 kpc and 794$\pm$23 kpc
respectively
\cite{LeeKimSarajedini2002,McConnachieIrwinFerguson2004}. It also
agrees with a geometric distance estimate of 800$\pm$180 kpc 
\cite{ArgonGreenhillMoran2004}.


The observed proper motion ${\vec v}_{prop}$ of a maser (e.g.,
M33/19 or IC\,133) in M33 can be decomposed into three components
${\vec v}_{prop}=  \vec v_{rot} + \vec v_\odot + \vec v_{M33}$.
Here $\vec v_{rot}$ is the motion of the maser due to the internal
galactic rotation in M33 and  $\vec v_\odot$ is the apparent motion of
M33 caused by the rotation of the Sun around the Galactic center. The last
contribution, $\vec v_{M33}$, is the proper motion of M33 relative
to the Milky Way.

The motion of the Sun can be decomposed into a circular
motion of the local standard of rest (LSR) and the peculiar motion of
the Sun. The peculiar motion of the Sun has been determined from
Hipparcos data \cite{DehnenBinney1998} to be U$_0$=10.00$\pm$0.36
km~s$^{-1}$ (radially inwards), V$_0$=5.25$\pm$0.62 km~s$^{-1}$ (in the
direction of Galactic rotation) and W$_0$=7.17$\pm$0.38 km~s$^{-1}$
(vertically upwards). The IAU adopted LSR moves with a velocity of 220
km s$^{-1}$ towards a Galactic longitude of $l=90^\circ$ and latitude of
$b=0^\circ$~\cite{KerrLynden-Bell1986}. New VLBI measurements of the
proper motion of Sgr A*, the compact radio source at the Galactic
center, indicate a slightly higher circular velocity of the LSR of
236$\pm$15 km~s$^{-1}$, for a distance of the Sun from the Galactic
center R$_0$=8 kpc, where the uncertainty is dominated by the
uncertainty in the distance to the Galactic center
\cite{ReidBrunthaler2004}. Using a LSR velocity of 236$\pm$15
km~s$^{-1}$ plus the peculiar velocity of the Sun from
\cite{DehnenBinney1998}, the motion of the Sun causes an apparent
proper motion of $\dot\alpha_{\odot}=52.5\pm3.3~\mu$as~year$^{-1}$ in
right ascension and $\dot\delta_{\odot}=-37.7\pm2.4~\mu$as~year$^{-1}$
in declination, assuming a distance of 730 kpc and the Galactic
coordinates of M33 ($l=133.6^\circ$, $b=-31.3^\circ$).   

Using the rotation model of \cite{CorbelliSchneider1997}, the
contribution from the rotation of M33 (for IC\,133) is
$\dot\alpha_{rot}=-18.5\pm6~\mu$as~year$^{-1}$  in right ascension and
$\dot\delta_{rot}=-21.6\pm6~\mu$as~year$^{-1}$ in declination. Here, we
assumed again an uncertainty of 20 km~s$^{-1}$ for the rotation
velocity and a distance of 730 kpc. Combining these velocity vectors,
we get the proper motion of M33: 

\begin{eqnarray}
\nonumber
\dot\alpha_{M33}=&\dot{\alpha}_{prop}-\dot\alpha_{rot}-\dot\alpha_\odot\\
\nonumber
=&(4.7\pm3.2 + 18.5\pm6 - 52.5\pm3.3) \frac{\mathrm{\mu as}} {\mathrm{year}}\\\nonumber
=&-29.3\pm7.6 \frac{\mathrm{\mu as}} {\mathrm{year}}=-101\pm35 \frac{km}{s}\\\nonumber
\mathrm{and}\\\nonumber
\dot\delta_{M33}=&\dot{\delta}_{prop}-\dot\delta_{rot}-\dot\delta_\odot\\\nonumber
=&(-14.1\pm6.4 +21.6\pm6 + 37.7\pm2.4) \frac{\mathrm{\mu as}} {\mathrm{year}}\\\nonumber
=&45.2\pm9.1 \frac{\mathrm{\mu as}} {\mathrm{year}}=156\pm47 \frac{km}{s}.
\end{eqnarray}

The transverse velocity changes by less than 5 km~s$^{-1}$ if we use
the TRGB distance of 794$\pm$23 kpc
\cite{McConnachieIrwinFerguson2004} for this analysis. Finally, the
systemic radial velocity of M33 is $-$179 km~s$^{-1}$
\cite{CorbelliSchneider1997}. The radial component of the rotation of
the Milky Way towards M33 is $-$140 $\pm$ 9 km~s$^{-1}$.  Hence, M33 is
moving with $-$39 $\pm$ 9 km~s$^{-1}$ towards the Milky Way. This gives
now the three dimensional velocity vector of M33 ({\it Fig.4}). The
total velocity of M33 relative to the Milky Way is 190 $\pm$ 59
km~s$^{-1}$.  


For Andromeda (M31), only one component of the three-dimensional
velocity vectoris known, the radial velocity of 116 km~s$^{-1}$ (301
km~s$^{-1}$ systemic velocity minus 185 km~s$^{-1}$ contribution from solar
motion) towards the Milky Way. However, the Milky Way is
possibly falling towards M31, because there are no
other large galaxies in the Local Group to generate angular momentum
through tidal torques \cite{KahnWoltjer1959}. Following this argument,
we assume a proper motion of 0 for M31. The geometry of the Andromeda
subgroup depends on the relative distance between M31 and M33. Thus, it
is crucial to use distances for the two galaxies that have similar
systematic errors, and we used the TRGB distances to M33 and M31
\cite{McConnachieIrwinFerguson2004,McConnachieIrwinFerguson2004b}.
Then the angle between the velocity vector of M33 relative to M31 and
the vector pointing from M33 towards M31 was 30$^\circ\pm$15$^\circ$. For
an angle of 30$^\circ$, only elliptical orbits with eccentricities of
e$>$0.88 are allowed. For the largest allowed angle of 45$^\circ$, the
eccentricities are restricted to e$>$0.7. This eccentricity limit
weakens if the proper motion of M31 is non-negligible and, for a motion
of $>$150 km~s$^{-1}$, any eccentricity is allowed.

If M33 is bound to M31, then the relative velocity of the two galaxies
must be smaller than the escape velocity. This gives -- for a zero
proper motion of M31 -- a lower limit for the mass of M31 of
1.2$\times$10$^{12}$ M$_\odot$. A substantial proper motion of M31
could reduce or increase the relative velocity and the lower mass limit
of M31. On the other hand, the dynamical friction of M31 on M33 indicates that
M31 cannot have a very massive halo of more than $\sim$10$^{12}$
M$_\odot$ unless the orbit of M33 has a low eccentricity. Otherwise,
the dynamical friction would have led to a decay of the orbit of M33
\cite{GottesmanHunterBoonyasait2002}. This agrees with a recent
estimates of $12.3^{+18}_{-6}\times10^{11}$~M$_\odot$ derived from the
three-dimensional positions and radial velocities of its satellite
galaxies \cite{EvansWilkinson2000}.

More than 80 years after van Maanen's observation, we have measured the
rotation and proper motion of M33. These measurements provide a method
to determine dynamical models for the Local Group and the mass and dark
matter halo of Andromeda and the Milky Way.    




\bibliography{1108342refs,scibib}

\begin{thebibliography}{10}
\bibitem{LG}
The Local Group is a small galaxy group in the local universe that includes the
  Milky Way, M31, M33 and 30 -- 40 other small galaxies within a radius of
  $\sim$ 1 Mpc.
\end{thebibliography}

\bibliographystyle{Science}


\begin{scilastnote}
\item The VLBA is operated by the National Radio Astronomy Observatory (NRAO).
The National Radio Astronomy Observatory is a facility of the National
Science Foundation operated under cooperative agreement by Associated
Universities, Inc. 
\end{scilastnote}


\clearpage

\begin{figure}
\includegraphics[width=12cm,angle=0]{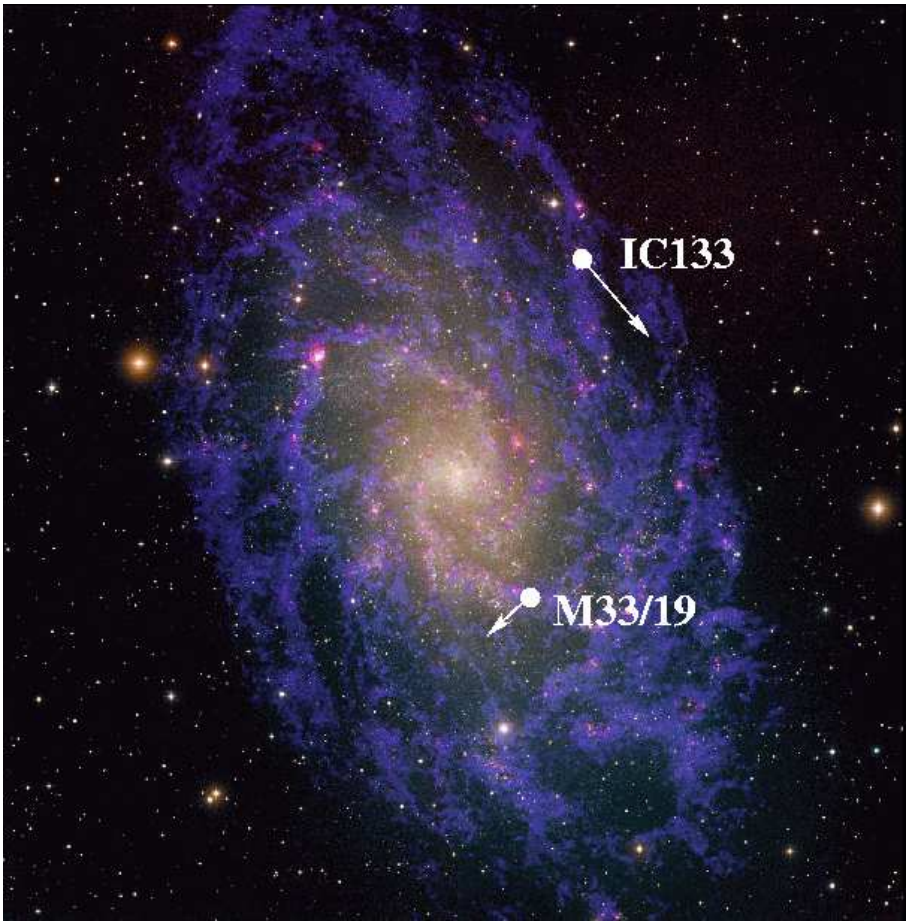}
\caption{The positions of two regions of maser activity in
  M33. Predicted motions due to rotation of the H\,I disk are also
  shown. Image courtesy of Travis Rector (NRAO/AUI/NSF and
  NOAO/AURA/NSF), David Thilker (NRAO/AUI/NSF) and Robert Braun
  (ASTRON).} 
\label{m33_pred}
\end{figure}

\begin{figure}
\includegraphics[width=12cm,angle=0]{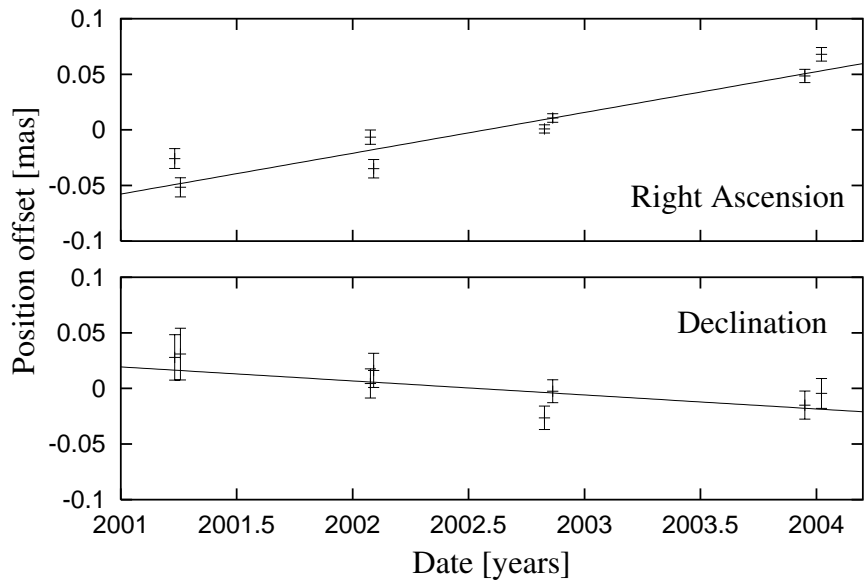}
\caption{Average position of the maser M33/19 in (top) right ascension and
  (bottom) declination relative to a background source.}
\label{m33_19}
\end{figure}
\begin{figure}
\includegraphics[width=12cm,angle=0]{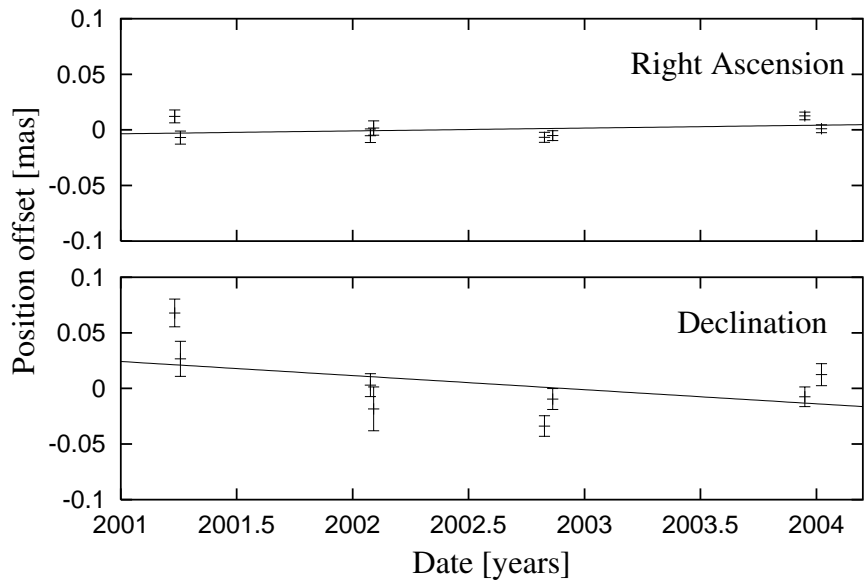}
\caption{Average position of the masers in (top) IC\,133 in right ascension
  and (bottom) declination relative to a background source.}
\label{ic133}
\end{figure}

\begin{figure}
\includegraphics[width=14cm,angle=0]{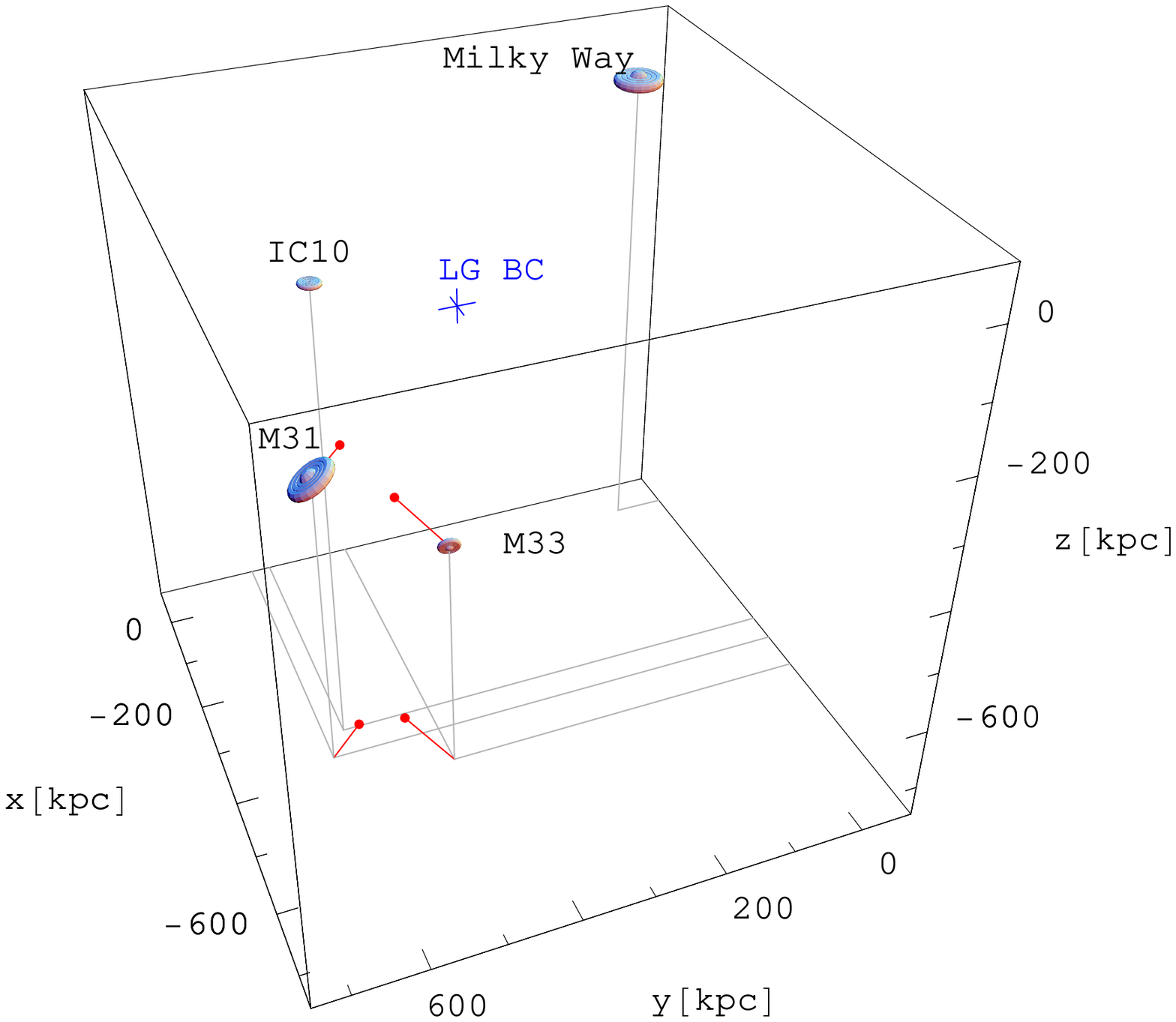}
\caption{Schematic view of the Local Group with the space velocity of
  M33 and the radial velocity of M31. The blue cross marks the
  position of the Local Group Barycenter (LG BC)
  \protect\cite{vandenBergh1999}.} 
\label{m33_mot}
\end{figure}





\end{document}